# Silicon Nitride resistance switching MIS cells doped with Silicon atoms

A. Mavropoulis [a], N. Vasileiadis [a,b], C. Bonafos [c], P. Normand [a], V. Ioannou-Sougleridis [a], G. Ch. Sirakoulis [b], P. Dimitrakis [a]

[a] *Institute of Nanoscience and Nanotechnology, NCSR "Demokritos", Ag. Paraskevi 15341, Greece*
[b] *Department of Electrical and Computer Engineering, Democritus University of Thrace, Xanthi 67100, Greece*
[c] *CEMES-CNRS et Université de Toulouse, BP94347-31055 Toulouse, Cedex 4, France*

A B S T R A C T

Stoichiometric $SiN_x$ layers (x=[N]/[Si]=1.33) are doped with Si atoms by ultra-low energy ion implantation (ULE-II) and subsequently annealed at different temperatures in inert ambient conditions. Detailed material and memory cells characterization is performed to investigate the effect of Si dopants on the switching properties and performance of the fabricated resistive memory cells. In this context extensive dc current-voltage and impedance spectroscopy measurements are carried out systematically and the role of doping in dielectric properties of the nitride films is enlightened. The dc and ac conduction mechanisms are investigated in a comprehensive way. Room temperature retention characteristics of resistive states are also presented.

## 1. Introduction

Silicon nitride ($Si_3N_4$) is the key material for charge-trapping (CT) nonvolatile memories for more than half of century [1, 2] in Flash SONOS [3] and Vertical CT-NVM technologies [4]. The nitrogen deficiency in the amorphous $SiN_x$ layers, due to the thermodynamics of the deposition methods and conditions, causes the formation of intrinsic defects with different configurations, i.e., nitrogen vacancies, Si-dangling bonds etc [5]. Recently, the CT-NVM technology was utilized to implement neuronal synapses. Furthermore, functional resistive memory (RRAM) cells with $SiN_x$ as active material have been demonstrated [6, 7]. The effect of nitride layer's stoichiometry x=[N]/[Si] (x=1.33 for $Si_3N_4$) as well as the nature of electrode material have been investigated in [6, 7] and [7, 8] respectively. Silicon nitride memristors exhibited very attractive properties for memory cell scaling [10], neuromorphic and in-memory and edge computing [11, 12, 13] as well as security applications [14, 15] realizing true-random number generators and physical unclonable functions. SiN-memristors can be programmed at different low-resistance states (LRS) well distinguished with respect to high-resistance state (HRS) exhibiting large operation window.

In addition, the role of doping in resistance switching dielectric materials for RRAMs has been investigated [16, 17]. In this work, nearly stoichiometric $SiN_x$ layers are doped with Si atoms by ultra-low energy ion implantation (ULE-II) and subsequent annealing at different temperatures in inert ambient. Detailed material and memory cells characterization is performed to investigate the effect of Si doping on the resistance switching properties and memory performance.

## 2. Experimental

LPCVD $SiN_x$ (x=1.27) 7nm layers were deposited on three different heavily Phosphorus-doped Si wafers ($N_d$=5×10$^{19}$ P/cm$^{-3}$) [17]. Then each wafer was implanted at 3keV with a different dose of Si-ions. A low temperature (low-*T*) and a high temperature (high-*T*) post-implantation annealing (PIA) were applied in order to heal ULE-II damage and incorporate Si atoms in SiN lattice, as shown in Table 1. A similar approach was followed in the past to form Si nanocrystals inside the nitride amorphous layer by ULE-II [18]. The structure of the examined samples is shown in Fig. 1. The calculated TRIM profiles of the implanted Si-atoms are shown in Fig.1 (inset). Top-electrode (TE) comprises 30nm Cu layer covered by 30nm Pt layer to prevent oxidation. For dc electrical characterization of the RRAM cells, HP4155 and Tektronix 4200A were used, while impedance measurements were carried out using HP4284 and Zurich Instruments MFIA.

**Table 1** Table of fabricated and examined samples

| Sample Id. | S12 | S13 | S22 | S23 | S42 | S43 |
|---|---|---|---|---|---|---|
| I.I. Dose (Si$^+$/cm$^{-3}$) | 1×10$^{14}$ | | 5×10$^{14}$ | | 5×10$^{15}$ | |
| Furnace Annealing | 800°C 20min N$_2$ | 950°C 20min N$_2$ | 800°C 20min N$_2$ | 950°C 20min N$_2$ | 800°C 20min N$_2$ | 950°C 20min N$_2$ |

## 3. Results and discussion

Standard *I-V* sweeps were performed under different current compliances, $I_{CC}$. Fig. 2 compares typical SET/RESET voltage sweeps from all samples at $I_{CC}$=100μA. For the sake of comparison, the *I-V* characteristic for the cell with stoichiometric film is presented. The current density of the Si-doped $SiN_x$ films before switching has changed significantly. Specifically, for the minimum and maximum implantation doses, the current before SET and after RESET, $I_{OFF}$, is significantly suppressed leading to higher SET and RESET voltages, $V_{SET}$ and $V_{RESET}$ respectively. This behavior is attributed to the annealing temperature. In Fig.3, statistical measurements of the $V_{SET}$ and $V_{RESET}$ are plotted. For comparison, the coefficient of variation $\sigma/\mu$, defined as the ratio of standard deviation $\sigma$ over the mean value $\mu$, is used. Evidently, for $V_{SET}$ the $\sigma/\mu$ is the same for any implantation dose and PIA temperature. However, $\sigma/\mu$ for $V_{RESET}$ of Si-doped samples annealed at low-*T* is higher than high-*T* annealed films suggesting that PIA at 800°C is partially healing the implantation damage. However, annealing at 950°C is more effective to heal the implantation defects as well as to allow implanted Si-atoms to react with the lattice atoms decreasing the cell-to-cell variation. All the examined MIS cells exhibited multi-level switching under variable $I_{CC}$ values. Typical, multilevel behavior is demonstrated in Fig. 4, denoting that the examined devices have prospects for multi-bit memory cells and in-memory computing. Obviously, increasing the implantation dose the LRS decreases allowing us to conclude that increasing the excess Si the formation of conductive filaments is easier. It should be stressed herein that the mechanism responsible for the resistance switching is related to the concentration of traps existing in $SiN_x$ films and specifically, the various forms of nitrogen vacancies and the resulting Si dangling bonds which under electric field form conductive filaments (CF) allowing the electrons to flow from one electrode to the other [19].

The conduction mechanisms in SiN thin films in memristive devices have been investigated [20]. The majority of the research investigations converge to the conclusion that the conduction mechanism is either trap-limited (space charge limited conduction, SCLC) [17, 8] or bulk-limited (Poole-Frenkel conduction, P-F) [10] or a combination of them (modified SCLC, MSCLC) [21]. The relations describing the above mechanisms are given by the following equations

$$I = \frac{9}{8} A\mu\varepsilon\varepsilon_0 \theta \frac{V^2}{d^3} \quad \text{where} \quad \theta = \frac{N_C}{N_T} e^{-q\varphi_T/k_B T} \qquad \text{for SCLC} \quad (1)$$

$$I = q\mu N_C A \left(\frac{V}{d}\right) exp\left[\frac{q(\varphi_T - \sqrt{qV/d\pi\varepsilon\varepsilon_0})}{k_B T}\right] = CVe^{\beta\sqrt{V}} \qquad \text{for P-F} \quad (2)$$

$$I = \alpha V^2 e^{\beta\sqrt{V}} \qquad \text{for MSCLC} \quad (3)$$

$$\alpha = \frac{9\varepsilon\varepsilon_0 \mu N_C A}{8d^3 N_T} e^{-q\varphi_T/k_B T} \qquad (3a)$$

$$\beta = \frac{q}{k_B T} \sqrt{\frac{q}{\pi\varepsilon\varepsilon_0 d}} \qquad (3b)$$

where $A$, $d$, $\varepsilon$, $N_C$, $\mu$ are the device area, the film thickness and dielectric constant, the density of states in conduction band of dielectric and the electronic drift mobility respectively and all the other symbols have their usual meaning. Also, $\varphi_T$ and $N_T$ are the trap energy and concentration respectively.

Figure 5 shows the analysis of *I-V* characteristics following P-F and MSCLC of Fig. 2 at HRS, before SET. Evidently both mechanisms fit very well. Nevertheless, P-F provides slightly better fitting coefficients. For non-implanted silicon nitride films (reference sample) the mechanism remains SCLC as has already been published [17]. This is a significant difference and is mainly attributed to the redistribution and/or modification of traps into the nitride film after ULE-II and PIA. The following intrinsic defects have been considered to exist in SiN: (i) a N vacancy, $V_N$, consisting of a $N_3\equiv Si\bullet$ and a $N_3\equiv Si-Si\equiv N_3$ adjacent units; (ii) a N vacancy, $V_N(H)$, saturated with H ($N_3\equiv Si-H$ and $N_3\equiv Si-Si\equiv N_3$); and (iii) a Si atom substitutional to N, Si—N, where a Si atom is bound to three other silicon atoms, $Si_3\equiv Si\bullet$ [5, 22]. According to equation (2) $q\varphi_T$ is the charge trap energy barrier height – the amount of energy required for an electron to escape the trap – and decreases with decreasing *x*, i.e., increasing the Si content in nitride films [23]. Once an electric field is applied, the trap depth is reduced. A trapped electron can then be thermally emitted with an increased probability over the reduced barrier into the conduction band of the dielectric. Following the conventional notation, $Si_3(\circ) / Si_3(^+)$ dangling bonds (filled/empty) are donor-like traps and will outnumber $N_2(\text{-}) / N_2(\circ)$ (filled/empty) acceptor-like states and hence the Fermi level will be pinned at $Si_3$ [24]. This is the case for implanted (not shown here) and non-implanted samples as has been already published [17]. For the Poole-Frenkel effect to occur, a trap must be positively charged when empty and neutral when filled [25, 26]. Thus, the Si dangling bonds formed due to Nitrogen ions movement towards the TE under electric field, exactly like Oxygen ions in valence change memories (VCM), are responsible for the creation of CF causing the resistance switching.

**Table 2** P-F linear fitting slope and calculated dielectric constants

| Sample | $\beta$ ($V^{-1/2}$) | $\varepsilon$ | Corrected $\varepsilon$ |
|---|---|---|---|
| S12 | 11.57 ± 0.05 | 10.07 ± 0.08 | 5.04 ± 0.04 |
| S22 | 10.93 ± 0.02 | 11.29 ± 0.04 | 5.65 ± 0.02 |
| S42 | 12.45 ± 0.06 | 8.70 ± 0.08 | 4.35 ± 0.04 |
| S13 | 10.14 ± 0.02 | 13.12 ± 0.05 | 6.56 ± 0.03 |
| S23 | 10.42 ± 0.02 | 12.42 ± 0.04 | 6.21 ± 0.02 |
| S43 | 11.40 ± 0.03 | 10.38 ± 0.05 | 5.19 ± 0.03 |

P-F plots shown in Fig.5 linearize equation (2) allowing for the calculation of the pre-exponential factor *C* from the intercept and the dielectric constant $\varepsilon$ from the slope $\beta$:

$$C = \frac{q\mu N_C A}{d} exp\left[\frac{q\varphi_T}{k_B T}\right] \qquad (4)$$

$$\varepsilon = \frac{q^3}{\pi\varepsilon_0 d} \frac{1}{(\beta k_B T)^2} \qquad (5)$$

The extraction of trap energy and electronic drift mobility in silicon nitride film requires measurements at different temperatures. It should be emphasized that *C* increases with trap density: the more silicon-rich the $SiN_x$ films become, the more the conductivity and the value of *C* increases. Nevertheless, the large difference in *C* is most likely due to an increase in the electronic drift mobility as the films become more silicon-rich [23]. The increase in *C* coincides with the increase in the current density for an increasing $\varepsilon$. Table 2 presents the calculated slope in P-F plots of Fig.5 and the extracted dielectric constant. Obviously, the values of $\varepsilon$ do not coincide with those in the literature for silicon nitride films, in the range 5 – 7.5.

**Table 3** Modeling parameters for Nyquist plots at LRS

| Sample | S12 | S22 | S42 | S13 | S23 | S43 | Ref. |
|---|---|---|---|---|---|---|---|
| $R_p$ (MΩ) | 1.02 | 0.39 | 0.51 | 0.54 | 0.83 | 0.59 | 0.008 |
| $C_p$ (pF) | 60.4 | 57.6 | 62.5 | 57.3 | 57.9 | 55.0 | 72.5 |
| $\varepsilon'$ @ 1kHz | 4.95 | 4.66 | 5.07 | 4.57 | 4.62 | 4.42 | 5.82 |

Further investigations include impedance spectroscopy at SET (LRS) and RESET (HRS) states for all devices listed in Table 3. Fig.6 shows the Nyquist plots at LRS, after voltage sweep and $I_{CC}$=100μA using an ac small signal 25mV. The results allow us to measure the exact value of the LRS and suggest very low series resistance $R_s$. The semicircle shape of the plots is indicative for the response of an $R_s$—$(R_p||C_p)$ circuit where $R_p$ is the resistance of the conductive paths in Si doped-SiN$_x$ and $C_p$ is the capacitive response of the remaining (unswitched) insulating volume in the examined MIS capacitor [27, 28]. Moreover, the analyses of the Nyquist plots (Fig.6) revealed the dependence of dielectric constant $\varepsilon'$ = Re($\varepsilon^*$) = $\varepsilon$ and ac conductance $\sigma'$=Re($\sigma^*$) for the investigated nitride films.

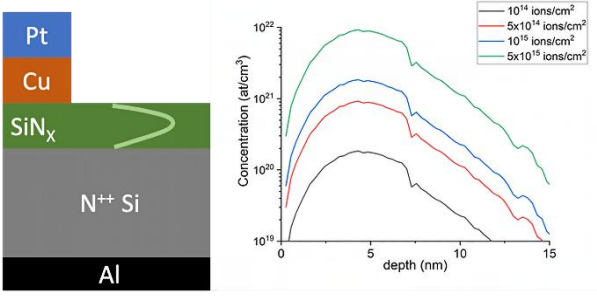

**Fig. 1.** Device structure description. Inset: TRIM calculation of the ULI-II Si distribution profile in the as-implanted SiN$_x$

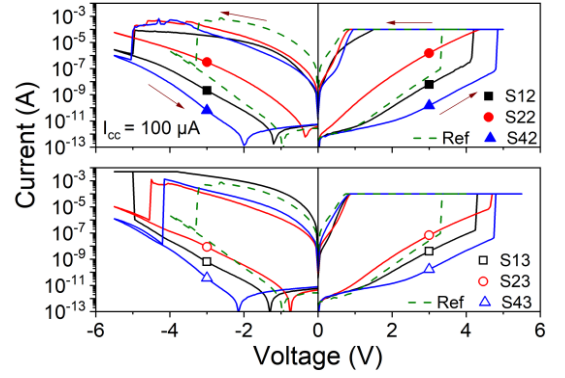

**Fig. 2.** Comparative plots of *I-V* sweeps for all the fabricated samples.

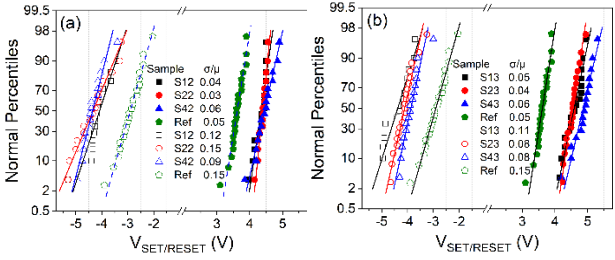

**Fig. 3.** Statistical distribution of $V_{SET}$ (solid symbols) and $V_{RESET}$ (hollow symbols) for (a) low-*T* and (b) high-*T* PIA ($I_{CC}$=100μA)

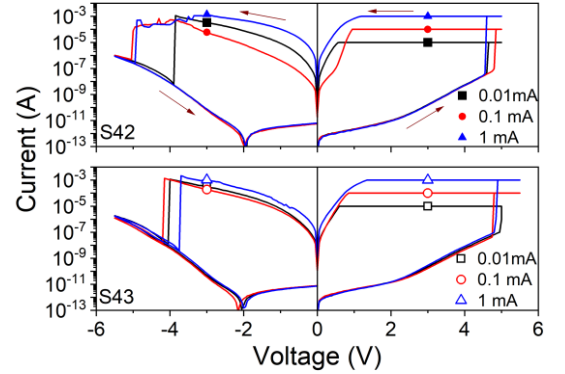

**Fig. 4.** Typical multi-level switching *I-V* curves for samples high-dose implanted samples at different $I_{CC}$

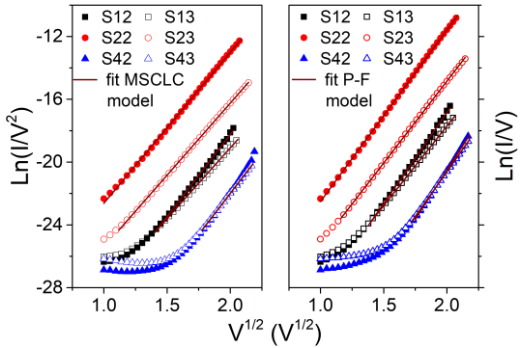

**Fig. 5.** P-F and MSCLC plots of I-V curves (HRS only) shown in Fig.4

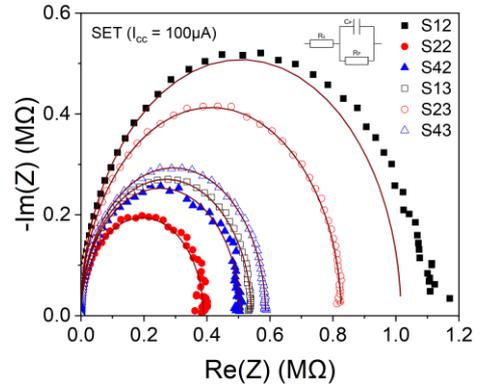

**Fig. 6.** Nyquist plots for the tested MIS cells at LRS. Inset: the equivalent circuit used to fit the data

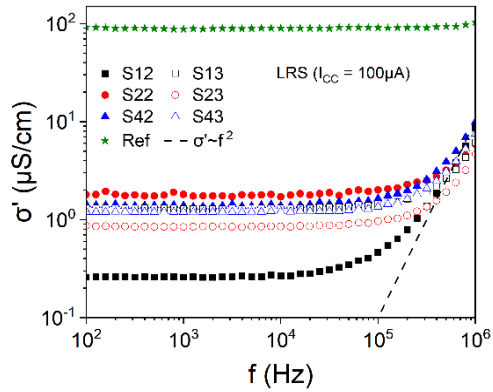
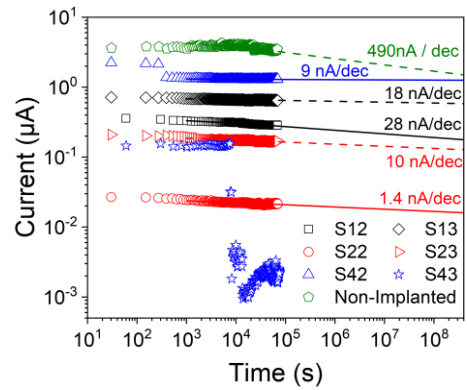

**Fig. 7.** Total real conductivity vs frequency measurements  **Fig. 8.** Room temperature retention measurements

Comparing the values of ε in tables 2 and 3 we found that impedance spectroscopy measurements revealed the correct values for SiN$_x$ films. The large values of ε calculated from P-F plots originate from the compensation of donor-traps by acceptor-traps [25], that is known also as "anomalous P-F effect" [26]. When there are acceptor sites present the intercept and slope of the ln($I/V$) vs. $V^{1/2}$ plot will change, because of the change of Fermi level with respect the defect/trapping levels, suggesting that both $C$ and ε are affected. Therefore, an acceptor compensation factor should be used to accommodate for this change [25, 26]. Thus, the original equation (2) is modified

$$I = q\mu N_C A \left(\frac{V}{d}\right) exp\left[\frac{q(\varphi_T - \sqrt{qV/d\pi\varepsilon\varepsilon_0})}{\xi k_B T}\right] \quad (6)$$

The compensation factor ξ is added in denominator of the exponential, where ξ=2 when there are no acceptor traps and ξ=1 when there is a considerable number of acceptor traps. The corrected values for ε are shown in Table 2, which are closer to the values obtained by impedance spectroscopy. The small variation is due to use of the same film thickness $d$, which has been proven that is affected with the incorporation of Si dopants by ULE-II (swelling effect) [18]. XTEM measurements are in progress in order to find the accurate value of $d$ for each sample. In general, ξ, can range between 1 and 2 depending on the position of the Fermi level [25, 26].

In Fig.7, typical ac conductance experimental measurements obtained at LRS are presented. The conductance $\sigma^*(f)$ is given by

$$\sigma^* = Re(\sigma^*) + Im(\sigma^*) = \sigma' + j\sigma'' = \sigma_{dc} + \sigma'_{ac} + j\sigma'' \quad (7)$$

The results suggest that for high-$T$ annealed layers the total real conductance σ' varies as ~$f^s$, where $s$ is close to 1.3 and 1.62 for LRS and HRS respectively. According to [29], the above values of $s$ denote that during SET and RESET the conduction in doped-SiN$_x$ layers consist of electron variable range hopping ($s$ is close to 1) and trap-to-trap tunneling mechanisms ($s$ is close to 2) respectively.

Room temperature (RT) retention measurements were also carried out for LRS defined by voltage sweep $I_{CC}$=100μA and the current decay rate was calculated, as shown in Fig. 8. By extrapolating the experimental results it is revealed that all samples retain their information for ten years except that implanted at highest dose with high-$T$ PIA.

## 4. Conclusions

The switching characteristics of ULE-II Si-doped SiN$_x$ thin layers were investigated. The conduction process is mainly governed by donor-like traps caused from the large density of trivalent Si-dangling bonds or similar defects. ac conductance measurements allow the identification of the conduction mechanism at LRS and HRS respectively. Ten years memory retention is achieved at RT.

**Declaration of Competing Interest**
The authors declare the following financial interests/personal relationships which may be considered as potential competing interests: This work was supported in part by the research projects "3D-TOPOS" (MIS 5131411) Operational Programme NSRF 2014-2020, General Secretariat of Research and Innovation (GSRI), Ministry of Development and "LIMA-chip" (Proj.No. 2748) which is funded by the Hellenic Foundation of Research and Innovation (HFRI).

**Acknowledgements**
This work was supported in part by the research projects "3D- TOPOS" (MIS 5131411) and "LIMA-chip" (Proj.No. 2748) which are funded by the Operational Programme NSRF 2014-2020 and the Hellenic Foundation of Research and Innovation (HFRI) respectively.